\documentclass[prl,twocolumn,aps,psfig,nofootinbib,nobibnotes,superscriptaddress]{revtex4}
\usepackage{graphicx}
\usepackage{bm,html}
\usepackage{color}
\newcommand{\J}{\,{\rm J}}

\newcommand{\etal}{{\em et al}}

\begin{document}

\title{{\sc Viewpoint}: A New Twist in Simulating Solar Flares}
\author{Axel Brandenburg}
\affiliation{Laboratory for Atmospheric and Space Physics,
JILA, and Department of Astrophysical and \\ Planetary Sciences,
Box 440, University of Colorado, Boulder, CO 80303, USA}
\affiliation{Nordita,
KTH Royal Institute of Technology and Stockholm University,
and \\ Department of Astronomy, Stockholm University, 10691 Stockholm, Sweden}

\date{\today}

\begin{abstract}
\noindent
Simulations show for the first time how the magnetic fields that produce
solar flares can extend out of the Sun by acquiring a twist.\footnote{
Viewpoint article on the paper by Chatterjee, Hansteen, \& Carlsson [1],
to be published in {\sc Physics}. Edited by Jessica Thomas.}
\end{abstract}

\maketitle

Protecting humans and their technologies from the harmful effects
of space weather is nothing new. Yet a recent event demonstrated how
fateful such events can be. On November 4, 2015, an active region on the
Sun known as NOAA 12443 unleashed a flare---a gust of high-speed charged
particles---whose radiation deactivated the radars at Sweden's
airports,\footnote[2]{\tiny\url{
http://www.thelocal.se/20151104/solar-storm-grounds-swedish-air-traffic}}
grounding all of the country's commercial flights. Such events underscore
our need to be able to predict when the next solar flare will occur and
how much energy it might release. This longstanding goal of astrophysics
will rely heavily on numerical simulations. Now, Piyali Chatterjee
of the University of Oslo, Norway, and her collaborators have taken
a significant step towards improving such simulations by developing
the first numerical model that explains how magnetic fields from
below the Sun's surface can produce energetic flares in the corona
[1]. Researchers have assumed that this was the mechanism for flares,
but it has been difficult to demonstrate with simulations. Chatterjee et
al.\ do so by making very simple, yet physically realistic, assumptions.

\begin{figure}[t!]\begin{center}
\includegraphics[width=\columnwidth]{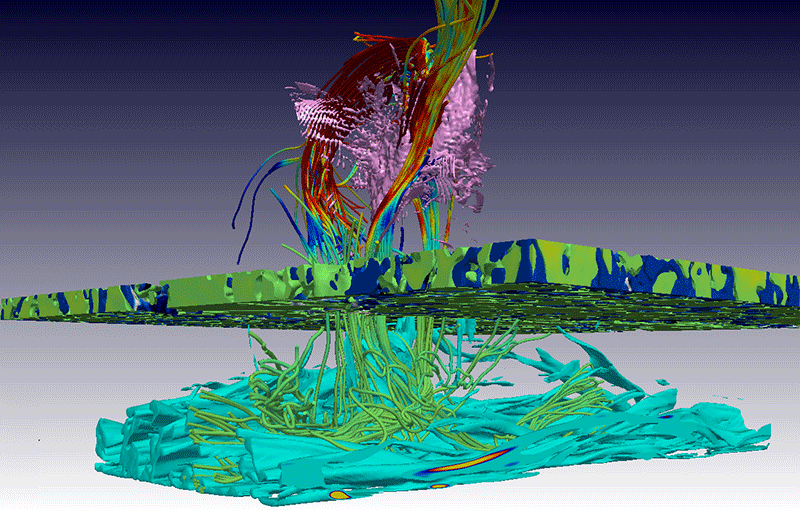}
\end{center}\caption[]{
Simulating a solar flare. The magnetic fields in Chatterjee et
al.'s simulations begin as a flat sheet of magnetized plasma (shown
in light blue). This sheet of magnetic flux breaks into tubes of flux,
which rise to the surface (dark blue and green slab), releasing energy at
specific points (the flare). Each flare is associated with a high-speed
jet of charged particles that squirts outward between reconnecting field
lines followed by a moderate decline in magnetic energy. The color of
a flux tube indicates the plasma velocity along the tube: red means upwards,
blue downwards, and green tubes have plasma velocities close to zero.
(P.\ Chatterjee/Univ.\ of Oslo)
}\label{ribbon}\end{figure}

Solar flares occur when giant ``tubes'' of magnetic flux in the Sun's
outer envelope, or corona, become unstable, releasing their stored
energy in an explosive burst. The magnitude of this energy is around
$10^{23}\J$, which is roughly equivalent to the energy delivered by the impact
of a 10-km-wide asteroid, such as the one causing the mass extinction
on Earth, 66 million years ago. Only a fraction of the flare's energy
strikes the Earth, but it can be enough to destroy electronics on Earth
and is a potential health risk to those who travel frequently by air,
particularly near the Arctic.

Although solar flares occur in the corona, the tubes of flux that give
rise to solar flares originate from the movement of ions and electrons
in the Sun's interior. In fact, it is widely believed that the tubes
form deep within the Sun, up to 200,000 km below its surface (the bottom
of the convection zone). What is unclear is how they are able to pass
through the Sun's turbulent interior and to its surface with enough
energy density left to produce a flare. One postulate is that the tubes
survive the ascent if their field lines twist around like the strands in
a braid. This twist provides an inward force that opposes the expansion
of the axial field during the tube's ascent.

So far, twist has been treated as a free parameter in simulations.
Chatterjee \etal.\ show that it can emerge naturally in
simulations that incorporate the Sun's rotation.
When a flux tube forms, the pressure associated with its field
displaces (or dilutes) plasma.
This makes the region within the tube lighter than
its surroundings, giving rise to an upwards
force---an effect known as magnetic buoyancy. As the tube
rises, the trajectories of its constituent ions ``swerve'' because
of the Sun's rotation (the Coriolis effect), causing the field
lines to become twisted.

\begin{figure}[t!]\begin{center}
\includegraphics[width=\columnwidth]{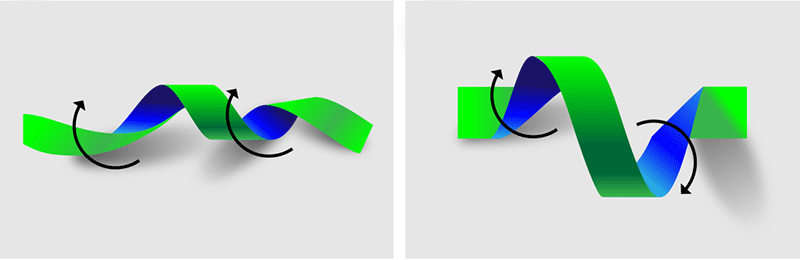}
\end{center}\caption[]{
Two types of twist with different helicities. At left, a
ribbon taped to a surface on both ends has a twist that only winds
in one sense (here, a right-handed twist). It has a net helicity
because the only way to remove the twist is to untape one of the
ends. At right, a ribbon has both a left-handed twist and a
right-handed twist. Its helicity is zero because the twist can be
undone without undoing either end.
(APS/Alan Stonebraker, \url{http://alanstonebraker.com/})
}\label{Fig2}\end{figure}

The researchers have recreated this process in their simulations.
They start with a sheet of magnetized plasma that
lies about 8000 km below the Sun’s surface and has a field
strength (parallel to the sheet) of ∼20,000 G. After 30 min
of solar time, the sheet breaks into (energetically more favorable)
separated tubes of flux, which become highly twisted by the
Sun's rotation by the time they reach the corona. The authors
show that flares, with energies of some $10^{23}\J$, occur
at localized sites where magnetic energy is converted into
kinetic energy and eventually into heat (see Fig.~1). These
events are associated with magnetic reconnection, where
field lines of opposite polarity break and reconnect. Chatterjee
\etal.\ also show that these flares can occur multiple
times when segments of the coronal field suddenly become
``overtwisted'' and release energy.

Up to now, simulations have initialized twist in an {\em ad hoc}
and unrealistic way; specifically, the twist only wound in
one sense (like a spiral staircase), as in the left side of Fig.~2.
However, in Chatterjee \etal.'s simulations, when a twist that
winds in one sense appears, it is always accompanied by a
counterwinding twist, as in the right side of Fig.~2. This conserves
what is known as the magnetic helicity, which is a
measure of twist integrated over volume. And it may be the
factor that helps the reconnection process associated with a flare,
because it increases the chance that a twisted field will be
surrounded by an oppositely wound field, allowing the two
to easily reconnect.

Several groups have simulated the emergence of magnetic
fields from below the convection zone into the corona.
But in many of these simulations, the magnetic fields were
untwisted. Most of the flux therefore ``accumulated'' just beneath
the surface without actually penetrating it [2], so no
flares were produced. But a twisted magnetic field, like that
emerging in Chatterjee \etal.'s simulations, wants to untwist.
It can do so much more easily once it is close to the surface,
where the surrounding pressure is smaller [3].

The advantage of the authors' approach is that the emergence
of twist is anchored in a physical phenomenon---the
Sun's rotation. However, the time it takes a flux tube to
rise to the surface because of magnetic buoyancy is only a
few hours, which is short compared with the Sun’s rotation
period of 25 days. One can therefore reasonably ask: How
could sufficient twist be generated in such a short time? The
answer might be that the Sun is highly stratified---the density
of plasma and its pressure change rapidly while moving
from the center to the surface. Rising plasma therefore expands
rapidly over a short distance, and since the Coriolis
effect on the plasma depends on the amount of expansion,
this enhances the effect of the Sun's rotation.

The simulations provide an estimate of the maximum flare
energy that the Sun, or another star, could potentially produce.
Chatterjee \etal.\ find that their model predicts flares
with energies as high at $10^{23}\J$, but the Kepler spacecraft has
detected many so-called superflares—flares producing energies
above $10^{26}\J$---from stars like our Sun [4]. Since there
is no tight correlation between flare frequency and stellar
rotation rate [5], superflares might be possible even for the
relatively slowly rotating Sun [6]. The consequences of superflares
for people and their technologies on the Sun-facing
side of the Earth are extreme, and future simulations based
on the model of Chatterjee \etal.\ might now be able to put
realistic limits on just how bad a worst-case scenario could
be.

\acknowledgments
A. Brandenburg acknowledges partial support from
the Swedish Research Council grant 2012-5797 and the
FRINATEK grant 231444 under the Research Council of Norway.

%r e f
\newcommand{\ybook}[3]{, {\em #2}. #3 (#1).}
\newcommand{\yanp}[3]{, Ann. Phys. {\bf #2}, #3 (#1).}
\newcommand{\yan}[3]{, Astron. Nachr. {\bf #2}, #3 (#1).}
\newcommand{\yapj}[3]{, Astrophys. J. {\bf #2}, #3 (#1).}
\newcommand{\yana}[3]{, Astron. Astrophys. {\bf #2}, #3 (#1).}
\newcommand{\ymn}[3]{, Mon.\ Not.\ R.\ Astron.\ Soc.\ {\bf #2}, #3 (#1).}
\newcommand{\ymnc}[3]{, Mon.\ Not.\ R.\ Astron.\ Soc.\ {\bf #2}, #3 (#1);}
\newcommand{\yjour}[4]{, #2 {\bf #3}, #4 (#1).}
\newcommand{\ynat}[3]{, Nature {\bf #2}, #3 (#1).}
\newcommand{\ypasj}[3]{, Publ. Astron. Soc. Japan {\bf #2}, #3 (#1).}
\newcommand{\yapjl}[3]{, Astrophys. J. Lett. {\bf #2}, #3 (#1).}
\newcommand{\papj}[2]{, Astrophys. J., in press, arXiv:#2  (#1).}
\newcommand{\sapj}[2]{, Astrophys. J., submitted, arXiv:#2  (#1).}
\newcommand{\sapjl}[2]{, Astrophys. J. Lett., submitted, arXiv:#2  (#1).}


\begin{thebibliography}{}

\bibitem{CHC}
P. Chatterjee, V. Hansteen, and M. Carlsson,
``Modeling Repeatedly Flaring $\delta$ Sunspots,''
Phys. Rev. Lett. 116, arXiv:1601.00749 (2016).

\bibitem{SAGT15}
P. Syntelis, V. Archontis, C. Gontikakis, and K. Tsinganos,
``Emergence of Non-Twisted Magnetic Fields in the Sun: Jets
and Atmospheric Response,'' Astron. Astrophys. 584, A10 (2015).

\bibitem{WBM12}
J. Warnecke, A. Brandenburg, and D. Mitra, ``Magnetic Twist:
A Source and Property of Space Weather,'' J. Space Weather
Space Clim. 2, A11 (2012).

\bibitem{Maehara}
H. Maehara, T. Shibayama, S. Notsu, Y. Notsu, T. Nagao, S.
Kusaba, S. Honda, D. Nogami, and K. Shibata, ``Superflares on
Solar-Type Stars,'' Nature 485, 478 (2012).

\bibitem{CHMBS14}
S. Candelaresi, A. Hillier, H. Maehara, A. Brandenburg, and K.
Shibata, ``Superflare Occurrence and Energies on G-, K-, and
M-type Dwarfs,'' Astrophys. J. 792, 67 (2014).

\bibitem{Shibata13}
K. Shibata \etal., ``Can Superflares Occur on Our Sun?'' Publ.
Astron. Soc. Jpn 65, 49 (2013).

\end{thebibliography}
\end{document}